\documentstyle[aps,prl,epsf,floats,twocolumn]{revtex}
\def\lap{\scriptsize${\stackrel{\textstyle _<}{_\sim}} $\normalsize \ }
\begin{document}
\draft
\twocolumn[\hsize\textwidth\columnwidth\hsize\csname @twocolumnfalse\endcsname
\title{Novel metallic behavior in two dimensions}
\author{X. G. Feng$^{1}$, Dragana Popovi\'{c}$^{1}$, S. 
Washburn$^{2}$, and V. Dobrosavljevi\'{c}$^{1}$}
\address{$^{1}$National High Magnetic Field Laboratory, 
Florida State University, Tallahassee, FL 32310 \\ 
$^{2}$Dept. of Physics and 
Astronomy, The University of North Carolina at Chapel Hill, Chapel Hill, NC 
27599}
\date{\today}
\maketitle

\begin{abstract}

Experiments on a sufficiently disordered two-dimensional (2D) electron system 
in silicon reveal a new and unexpected kind of metallic behavior, where the 
conductivity decreases as \linebreak 
$\sigma (n_s,T)=\sigma(n_s,T=0)+A(n_s)\, T^2$ ($n_s$
-- carrier density) to a {\em non-zero} value as temperature 
$T\rightarrow 0$.  In 2D, the 
existence of a metal with $d\sigma/dT > 0$ is very surprising.  In addition, a
novel type of a metal-insulator transition obtains, which is unlike any known 
quantum phase transition in 2D.  

\end{abstract}

]

The transport behavior indicative of the existence of a metallic
state at zero temperature in two dimensions (2D) has been observed in a 
variety of 2D electron~\cite{Krav,DP_MIT,electrons} and
hole~\cite{holes} systems.  However, there is still no
generally accepted microscopic description of the apparent metallic phase and 
the metal-insulator transition (MIT) in 2D.  Until now, the metallic phase 
has been characterized by an increase of conductivity $\sigma$ as temperature 
$T\rightarrow 0$.  
In the presence of a particular type of disorder, such a metallic phase is
suppressed~\cite{spinflip}.  Here we show that this 
gives rise to a new and unexpected kind of metallic behavior, 
where $\sigma$ decreases but {\em does not go to zero} (as expected for an 
insulator) when $T\rightarrow 0$.  While unambiguously established by our 
data, this behavior is in a striking contradiction with any theoretical 
description available to date.  In addition, we report the discovery of a 
novel type of a MIT, which occurs between this metal and an insulating state 
as the carrier density $n_s$ is reduced.  While the transition is analogous to
that observed in some three-dimensional (3D) systems, it is unlike any known 
quantum phase transition in 2D.  We point out that our samples are 
representative of a {\em broad} class of Si metal-oxide-semiconductor 
field-effect transistors (MOSFETs) historically known as ``nonideal'' 
samples~\cite{AFS}.

In Si MOSFETs, the 2D electron system is formed by confining the
electrons to the Si side of the Si-SiO$_2$ interface.  At low
$T$, electrons are quantized in the lowest energy level (subband)
for motion perpendicular to the interface, and are thus forced to move in a 
plane parallel to it.  $n_s$ is varied by applying 
voltage $V_g$ to the metallic gate.  As a result of disorder (potential 
scattering by the charges in the oxide and by the roughness of the interface),
the 2D density of states of each subband acquires a tail of strongly localized
states.  In {\em sufficiently disordered samples}, such as ours, the band 
tails may be so long that some of the localized states in the tail of the 
upper subband~\cite{sub_comment} may be populated even at low $n_s$, and act 
as additional scattering centers for 2D electrons~\cite{sub_scatt}.  In 
particular, since at least some of them must be singly populated due to a 
large on-site Coulomb repulsion
(tens of meV), it is plausible that they may act as local magnetic moments.  
We have shown earlier~\cite{spinflip} that an arbitrarily small amount of such
scattering suppresses the $d\sigma/dT <0$ behavior in the $T\rightarrow 0$ 
limit.  We note that, strictly speaking, since the carriers in the ground 
subband can exchange with the carriers in the tail of the upper subband, there
is some hybridization between the two but the important question is the time 
scale for the various different mechanisms ({\it e.~g.} for the exchange and 
for the mean free time for transport).  We also point out that some evidence
for qualitatively the same behavior of $\sigma (T)$ was found~\cite{spinflip} 
even in the case when {\em only} the ground subband was occupied but the 
potential scattering was increased considerably by applying large negative 
voltage ($V_{sub}$) to the Si substrate.  The results were interpreted as
being due to the formation of disorder-induced local moments (singly occupied
localized states) associated with the ground subband.  We realize, however,
that the detailed microscopic picture is probably more complex, as discussed 
below.

By varying $V_{sub}$, the splitting of the subbands and hence the population of
the tail of the upper subband are changed.  
In the presence of occupied ``upper tail'' states, 
in the metallic phase $d\sigma/dT$ changes 
sign from positive to negative at $T=T_m (n_s,V_{sub})$ as $T$ 
increases~\cite{spinflip}.  For $T > T_m$, we observe~\cite{DP_MIT,spinflip} 
the ``conventional'' 2D metallic behavior with $d\sigma/dT < 0$ and other 
features ({\it e.~g.} scaling) that reflect the ``conventional''~\cite{Krav} 
MIT.  The focus of this study, however, is the low-$T$ ($T<T_m$) regime.
We concentrate on the data obtained with $V_{sub}=+1$~V, because the reduced 
subband splitting leads to a large number of 
occupied ``upper tail'' states and $T_m > 4.5$~K, and hence to the observation
of the novel behavior of $\sigma$ up to relatively high $T$ ($\sim 2$~K).  
The same qualitative behavior is also 
observed when the number of occupied ``upper tail'' states
is smaller ({\it e.~g.} for negative
$V_{sub}$), the only difference being that the low-$T$ ($T < T_m$) 
regime becomes smaller~\cite{spinflip}.  For $V_{sub}=+1$~V, the number
of populated states in the tail of the upper subband
is roughly constant for all $n_s$ that are of interest here
($n_s$\lap $n_{max}\sim 5\times 10^{15}$m$^{-2}$, where $n_{max}$ is the 
density where mobility reaches its maximum)~\cite{dpmoriond99}.  

\begin{figure}[t]
\epsfxsize=3.2in \epsfbox{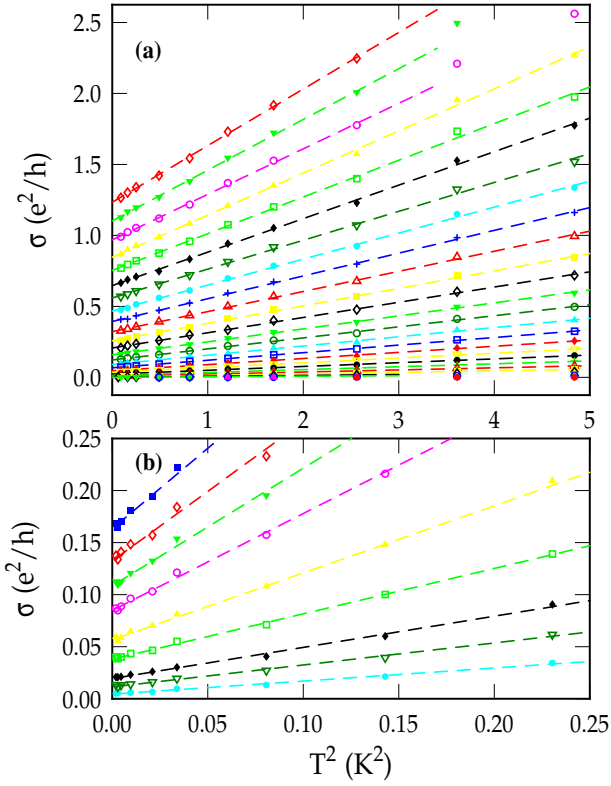}\vspace{5pt}
\caption{$\sigma (T)$ plotted {\it vs.}
$T^2$ for different $n_s$.  The dashed lines are fits.  (a) Sample
12: $n_s$ varies from $3.0\times 10^{15}$m$^{-2}$ (top) 
down to $0.7\times 10^{15}$m$^{-2}$ (bottom) in steps of $0.1\times 
10^{15}$m$^{-2}$ and $0.3\leq T < 2.2$~K.  (b) Sample 11: 
top to bottom, $n_s=3.8, 3.7, 3.6\times 10^{15}$m$^{-2}$, and $3.5\times
10^{15}$m$^{-2}$ to $2.5\times 10^{15}$m$^{-2}$ in steps of $0.2\times 
10^{15}$m$^{-2}$; $0.045\leq T < 0.5$~K.}
\vspace{-6pt}
\label{T2}
\end{figure}

Our samples are standard two-terminal Si MOSFETs of Corbino (circular) 
geometry (channel length $=0.4$~mm, mean circumference $=8$~mm), and a peak 
mobility $\sim 1$~m$^2$/Vs at 4.2~K.  Other sample details have been given 
elsewhere~\cite{DP_MIT,contacts}.  Conductance was measured as a function of 
$V_g$ using a low-noise current preamplifier and a low-noise analog lock-in at
$\sim 13$~Hz (the lead resistance was subtracted by the usual method).  The 
excitation voltage $V_{exc}$ was kept constant and low enough to avoid 
electron heating~\cite{heating}.
Most of the measurements were carried out in a $He^3$ cryostat 
(base $T=0.247$~K).  Figure~\ref{T2}(a) shows $\sigma (T)$ for different 
values of $n_s$.  An 
excellent fit (dashed lines) to the data is obtained with 
\begin{equation}
\sigma (n_s,T)=\sigma (n_s,T=0)+A(n_s)\, T^2
\label{T2eq}
\end{equation}
over a wide range of $n_s$ as shown, where $\sigma$ measured at the lowest $T$ 
(0.3~K) varies over three orders of magnitude below $e^2/h$ ($e^2/h$ -- 
quantum unit of conductance).  For each $n_s$, the best exponent in the power 
law dependence of $\sigma$ on $T$ [Eq.~(\ref{T2eq})] lies within 10\% of 2.  
Several samples have been studied and they all exhibit qualitatively the same 
behavior.  Fig.~\ref{T2}(b) shows the data obtained on another sample (\# 11) 
at $T$ down to 0.045~K.  
Obviously, Eq.~(\ref{T2eq}) describes the data well even at this, an order of 
magnitude lower $T$~\cite{heating1}.
At such low $T$, it is not possible to study the behavior very close to the MIT
because of the small signal-to-noise ratio.  The observed $\sigma (T)$
spans, therefore, two decades in $T$, starting from 
$T\sim 2$~K, and clearly cannot be explained by heating effects.

\begin{figure}[t]
\epsfxsize=3.2in \epsfbox{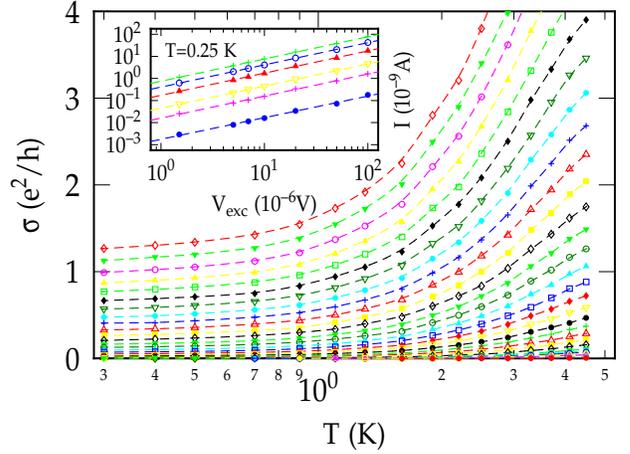}\vspace{5pt}
\caption{The same data as in Fig.~1(a) plotted {\it vs.} $\ln T$.
The dashed lines guide the eye.
Inset: Current $I$ (for sample 17) {\it vs.} $V_{exc}$ for 
$n_{s}(10^{15}$m$^{-2})=3.0, 2.5, 2.0, 1.5, 1.25, 1.0$ (top to bottom).
The dashed lines are fits with the slopes equal to 1.
}
\vspace{-6pt}
\label{lnT}
\end{figure}

In order to compare our data to the predictions of the theory of weak
localization~\cite{LR}, we plot the data from Fig.~\ref{T2}(a) {\it vs.}
$\ln T$.  Fig.~\ref{lnT} shows that $\sigma$ cannot be described by $\ln T$ in
any range of $T$ and, in fact, the curvature of the observed $\sigma (T)$ is
the {\em opposite} from the one expected for an insulating state.  We note 
that the conduction at 0.25~K was ohmic~\cite{ohmic} for $V_{exc}$ of up to at
least 100~$\mu$V (Fig.~\ref{lnT} inset) but $V_{exc}\sim 10~\mu$V was 
typically used at $T\geq 0.25$~K.

\begin{figure}[t]
\epsfxsize=3.2in \epsfbox{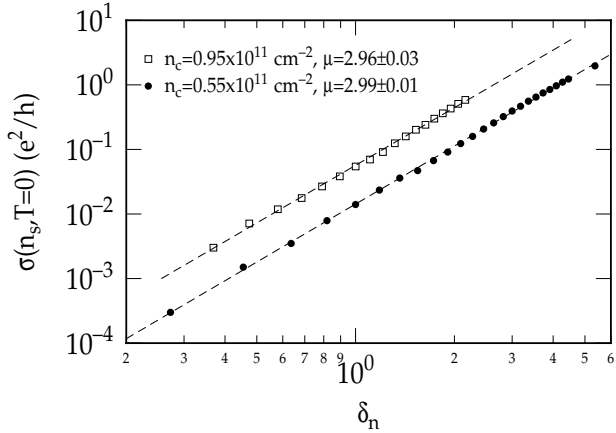}\vspace{5pt}
\caption{$\sigma (n_s,T=0)$ {\it vs.} the distance from the MIT for samples 9 
(squares) and 12 (dots).  The dashed lines are fits with the slopes equal to 
the critical exponent $\mu$.}
\vspace{-6pt}
\label{fig3}
\end{figure}

The high quality of the fits [Eq.~(\ref{T2eq})] allows a reliable 
extrapolation of $\sigma (n_s,T=0)$, whose
finite ({\it i.~e.} non-zero) values mean that, in spite of the decrease of 
$\sigma (n_s,T)$ with decreasing $T$, {\it the 2D system is in the metallic 
state}.  $\sigma(n_s,T=0)$ is shown in Fig.~\ref{fig3} for two different
samples as a function of $\delta_n = (n_s-n_c)/n_c$, the distance 
from the MIT ($n_c$ -- critical density).  For sample 12, for example, 
the Fermi energy $E_F\approx 4$~K at the MIT, and $r_s\approx 22$ ($r_s$ is 
the average inter-electron separation in units of the effective Bohr radius). 
We find the power-law behavior $\sigma (n_s,T=0)\sim\delta_{n}^{\mu}$ 
($\mu =2.99\pm 0.01$ for sample 12), as expected in the vicinity of a quantum 
critical point~\cite{Goldenfeld}, such as the MIT.  The power law holds over a
very wide range of $\delta_n$ (up to 5) similar to what has been 
observed~\cite{Rosenbaum} in Si:P near the MIT.  In addition, even though the
MIT occurs at different $n_c$ in different samples, the critical exponents 
$\mu$ are the same in both samples (Fig.~\ref{fig3}), as expected from general
arguments~\cite{Goldenfeld}.

\begin{figure}
\epsfxsize=3.2in \epsfbox{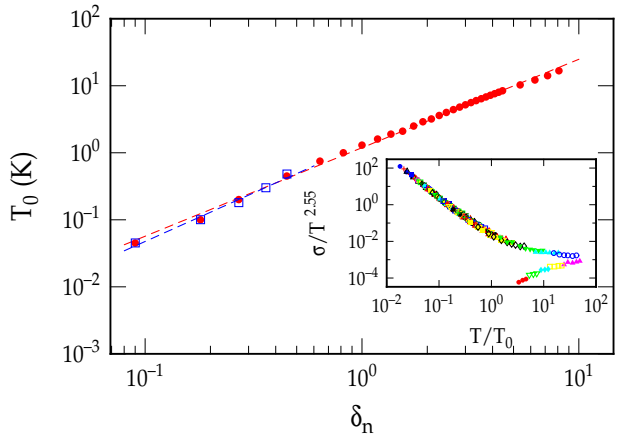}\vspace{5pt}
\caption{Scaling parameter $T_0$ as a function of $|\delta_n|$ for sample 12. 
Open symbols correspond to the insulating side of the transition, closed 
symbols to the metallic one.  The dashed lines are fits with slopes 
$1.4\pm 0.1$ and $1.32\pm 0.01$, respectively.  Inset, scaling of raw data 
$\sigma/\sigma_c\sim\sigma/T^x$ in units of $e^2/h$K$^{2.55}$.  Different 
symbols correspond to different $n_s$ ranging from $0.3\times 10^{15}$m$^{-2}$
to $3.0\times 10^{15}$m$^{-2}$.  It was possible to scale all the data below 
about 2~K.}
\vspace{-6pt}
\label{scaling}
\end{figure}

In addition, very general considerations have suggested~\cite{scaling_review}
that the conductivity near the MIT can be described by a scaling form 
\begin{equation}
\sigma(n_s,T) = \sigma_c (T)f(T/\delta_{n}^{z\nu}),
\label{eq}
\end{equation}
where $z$ and $\nu$ are the dynamical and correlation length exponents, 
respectively, and the critical conductivity 
\begin{equation}
\sigma_c = \sigma (n_s=n_c,T)\sim T^x.
\label{x}
\end{equation}
Indeed, we find such a power-law dependence, with $x=2.55\pm 0.3$ for sample 
12, and $x=2.7\pm 0.5$ for sample 9.  We show in the inset of 
Fig.~\ref{scaling} that,
for the given range of $n_s$ and $T$, all 
$\sigma (n_s,T)/\sigma_c (T)\sim\sigma (n_s,T)/T^x$ 
collapse onto a single scaling function $f(T/T_0)$.  $T_0$
is displayed in Fig.~\ref{scaling} as a function of $\delta_n$ on both sides 
of the transition.  Again, we find a power-law behavior, $T_0\sim 
|\delta_n|^{z\nu}$, in agreement with Eq.~(\ref{eq}).  For the metallic side 
of the transition, $z\nu = 1.32\pm 0.01$, and for the insulating side $z\nu = 
1.4\pm 0.1$.  This value of $z\nu$ is similar to $z\nu=0.8-1.7$ 
obtained~\cite{Krav,DP_MIT,spinflip,miscznu} in Si MOSFETs for the 2D
MIT that occurs between an insulator and a metal with $d\sigma/dT < 0$.  
We also note that Eq.~(\ref{T2eq}) describes the data well only in the low-$T$
metallic regime, {\it i.~e.} for $T<T_0(\delta_n)$.  In general, a different 
$\sigma (T)$ is expected in the quantum critical regime [$T>T_0(\delta_n)$], 
as given by Eq.~(\ref{x})~\cite{scaling_review}, and this is exactly what we 
find.  For example, for $0.3\leq T<2$~K, all data for $\delta_n$~\lap 0.25 in 
Fig.~\ref{T2}(a) follow Eq.~(\ref{x}) ({\it i.~e.} $\sigma (n_s,T)\sim T^x$).

It is easy to show using standard scaling 
arguments~\cite{scaling_review} that $\mu = x(z\nu)$, {\it i.~e.} that $\mu$ 
can be determined not only from extrapolations of $\sigma (T)$ to $T=0$ (see 
Eq.~(\ref{T2eq}) and Fig.~\ref{fig3}) but also based on all data taken at all 
$T$ and values of $n_s$ for which scaling holds.  In this way, we 
obtain $\mu = x(z\nu)=3.4\pm 0.4$ (for sample 12), in agreement with the 
method used above.  We note that the uncertainty in $\mu$ obtained from 
dynamical scaling is comparable in size to those of various critical exponents
in other systems ($\approx$5--30\%)~\cite{Myriam_review}.  By contrast, $\mu$ 
found from zero-temperature extrapolation of $\sigma (T)$ is determined with a
much higher accuracy, to within 0.3\% (Fig.~\ref{fig3}).  This is especially 
remarkable since $\sigma (n_s,T=0)$ (for sample 12) spans four orders of 
magnitude.

Several striking features of our data stand out. 

(1) $\sigma (T)$ follows a precise $T^2$ form over a very broad $T$ range. 
Such behavior is well established for metals 
containing local magnetic moments, and is believed to result from the Kondo 
effect~\cite{Hewson}.  In fact, to the best of our knowledge, there is no 
other known mechanism that results in an {\em increase} of $\sigma$ as 
$T^2$.  In our case, this feature provides the most direct 
evidence of the presence of local magnetic moments.  
It is worth noticing, however, that exchange with any kind of two-level systems
is capable of mimicking much of the Kondo effect~\cite{Vladar}.  Even though
there may be some such shallow traps present at the Si-SiO$_2$ interface, it is
difficult to rationalize the non-monotonic dependence on $V_{sub}$ that we 
observe~\cite{spinflip} based on any interface trap model.  If the two-level
systems are indeed responsible for the observed $\sigma (T)$, it is more likely
that they are related to the occupied states in the tail of the 
upper subband~\cite{local99}.  In general, one expects the $T^2$ behavior for a
quantum impurity embedded in a Fermi liquid in any dimension.  On the other
hand, it has been suggested~\cite{newgang} that the metallic state in 2D may 
not have a Fermi liquid character, but this interesting question requires
further theoretical and experimental work.  
In particular, careful experiments in a parallel magnetic field should provide
valuable insights into the microscopic nature of the conduction mechanism.  

We point out that the very existence of the MIT and the associated dependence 
of various quantities on $n_s$ near the critical region, reflect the 
importance of the localization effects in addition to the interaction 
mechanisms such as the Kondo effect.  From a practical point of view and 
regardless of the interpretation, such a simple $\sigma (T)$ 
[Eq.~(\ref{T2eq})] is crucial for characterizing the low-$T$ state of the 
system since it allows for an unambiguous extrapolation to $T=0$.  

(2) Scattering of the 2D conduction electrons by electrons populating the tail
of the upper subband leads to a qualitative modification of the low-$T$ 
transport, and to a novel metallic behavior characterized by $d\sigma/dT > 0$.
Our data demonstrate clearly that, contrary to what is usually assumed, the 
mere decrease of $\sigma$ with decreasing $T$ at a given $n_s$ {\em does not 
necessarily imply the existence of an insulating state at $T=0$}.  Hence, for
example, without a detailed study of $\sigma (T)$ and without any evidence for
scaling, the observed~\cite{reent} change from $d\sigma/dT < 0$ to
$d\sigma/dT >0$ at high $n_s$ does {\it not} represent an evidence for
the second MIT as $n_s$ is varied.

(3) In the 2D metal studied here, 
$\sigma (n_s,T=0)$ {\it decreases continuously}, and follows a distinct 
power-law behavior as the MIT is approached. In particular, {\it metallic} 
$\sigma$ as small as $10^{-3} e^2/h$ has been observed, 
in a striking contrast to anything that has been reported in other 2D systems
when $d\sigma/dT < 0$.  A similar observation in 3D 
systems~\cite{Rosenbaum} has demonstrated the absence of minimum metallic 
conductivity, and has had a profound impact on shaping the theoretical ideas 
about the MIT.

(4) Our data show an excellent fit to the dynamical scaling described by 
Eq.~(\ref{eq}). It should be noted that such scaling behavior is not consistent
with the simple single-parameter scaling hypothesis~\cite{newgang}.  In
particular, this formulation~\cite{newgang} allows only for a ``metallic-like''
temperature dependence ($d\sigma/dT < 0$) in the conducting phase, in contrast
to what we find.  In addition, the exponent $x$ in Eq.~(\ref{x}) is expected 
to take the value $x = (D-2)/z$, and thus to vanish in 2D.  On the
other hand, it should be emphasized that $x\neq 0$ does {\it not} 
contradict~\cite{scaling_review} any fundamental principle even for 2D 
systems. Indeed, such violations of ``Wegner scaling''~\cite{scaling_review} 
were predicted for certain microscopic models in presence of ``dangerously
irrelevant operators''~\cite{twoparam}.  Interestingly, recent work on 
the 2D MIT for a $d\sigma/dT < 0$ metal has found 
evidence~\cite{NamJung,local99} 
that the {\it same} scaling form can be used to resolve some apparent 
violations of the single-parameter scaling. In that case though, the exponent 
$x$ takes a distinctly different value, presumably reflecting the different 
universality classes of the two situations.

Our study shows that, in the presence of a particular kind of disorder,
a new type of metallic behavior and a MIT are 
exhibited in a 2D electron system.  In contrast to the 2D conducting phase and
the MIT discovered by Kravchenko {\it et al.}~\cite{Krav}, the 2D metal studied
here has $d\sigma/dT > 0$, and the scaling form that describes the 
conductivity near the MIT [Eq.~(\ref{eq})] includes a $T$-{\em 
dependent} prefactor $\sigma_c(T)$, analogous to the MIT in 3D
systems.

We are grateful to K. Eng for technical assistance.  This work was supported 
by NSF Grant DMR-9796339, NHMFL through NSF 
Cooperative Agreement DMR-9527035, an NHMFL In-House Research Program 
grant (D. P.), NSF Grant DMR-9974311 (V. D.), and Alfred P. Sloan Foundation 
(V. D.).


\begin{references}
\bibitem{sub_comment} Usually, this is a tail of the $E_{0'}$ subband -- the
lowest subband associated with the four conduction band valleys with the
light effective mass (0.190~$m_e$) perpendicular to the interface.  See also
Ref.~\cite{AFS}.
\bibitem{contacts} Measurements in the quantum Hall regime indicate that
contact resistances are less than a few Ohms, %in the $T$ range of interest.  
and the response to $V_{sub}$~\cite{spinflip} proves that the measured effects 
arise from the 2D channel -- not the contacts.
\bibitem{heating} For example, at the lowest bath $T$ (0.020~K), where 
the \linebreak sample was immersed in the mixing chamber of a $He^{3}-He^{4}$ dilution 
refrigerator, $V_{exc}\sim 1~\mu$V rms (eV=0.011~K $<kT$), so the total 
power dissipated in the sample was only between $10^{-16}$~W and $10^{-14}$~W,
depending on $n_s$.
\bibitem{ohmic} Amplitudes for any possible quadratic terms are statistically
insignificant.
\end{references}
\end{document}